\documentclass[fleqn,twoside]{article}
\usepackage{espcrc2}

\usepackage{graphicx}

\newcommand{\AmS}{{\protect\the\textfont2
  A\kern-.1667em\lower.5ex\hbox{M}\kern-.125emS}}

\title{Hadronic decays from the lattice}

\author{UKQCD Collaboration: C. Michael\address{Theoretical Physics Division, 
Dept. of Mathematical Sciences,  \\       
  University of Liverpool, Liverpool L69 3BX, UK }
 and C. McNeile\addressmark 
 }

\begin{document}

\begin{abstract}
 We discuss strategies to determine hadronic decay couplings from 
lattice studies. 
 As an application, we explore the decay of a vector meson to two
pseudoscalar mesons with $N_f=2$ flavours of sea quark.
Although we are working with quark masses that do not allow a physical
decay, we show how  the transition rate can be evaluated from the 
amplitude for $\rho \to \pi \pi$  and from the annihilation component of
$\pi \pi \to \pi \pi$. We explore  the decay amplitude for two different
pion momenta and find consistent results. The coupling strength we find
is in agreement with experiment.  We also find evidence for a shift in
the $\rho$ mass caused  by mixing with two pion states. 
\vspace{1pc}
\end{abstract}

\maketitle

\section{INTRODUCTION}

 The recent renaissance of experiments to study the spectroscopy of
light hadrons is partly driven by the goal to understand confinement.
Any analysis of experimental data requires both a calculation of decay
widths as well as the masses\cite{Close:2001zp}.  Although computing the
masses of resonances is (currently) not part of the program to compute
``gold plated''  observables with high precision from lattice QCD,
dealing with decay widths is an inherent part  of studying  hadrons
which are under current experimental scrutiny, such as ``exotics'', and
gluonic components of scalar mesons. Here we compute the  coupling of the
$\rho$ meson to two pions  to validate our procedure.

 One limitation of the  lattice approach  to QCD is  in exploring
hadronic decays because the  lattice, using Euclidean time, has
important contributions from low lying thresholds~\cite{cmdecay} which 
can obstruct the study of decay widths.  The finite spatial size of the
lattice implies that two-body  states are actually discrete. By
measuring their energy very precisely as the spatial  volume is varied,
it is possible~\cite{luscher} to extract the scattering phase shifts and
hence decay properties. For on-shell transitions,  it is  possible to
estimate hadronic transition strengths  more directly  and this approach
has been used to explore~\cite{hdecay} hybrid meson  decay rates. Here
we explore this approach further  for the case of  $\rho$ meson decay to
$\pi \pi$, following ref.~\cite{rhodecay}.

\section{TRANSITIONS}

\begin{figure}[th]
 \vspace{-3cm}  
  \includegraphics[scale=0.3]{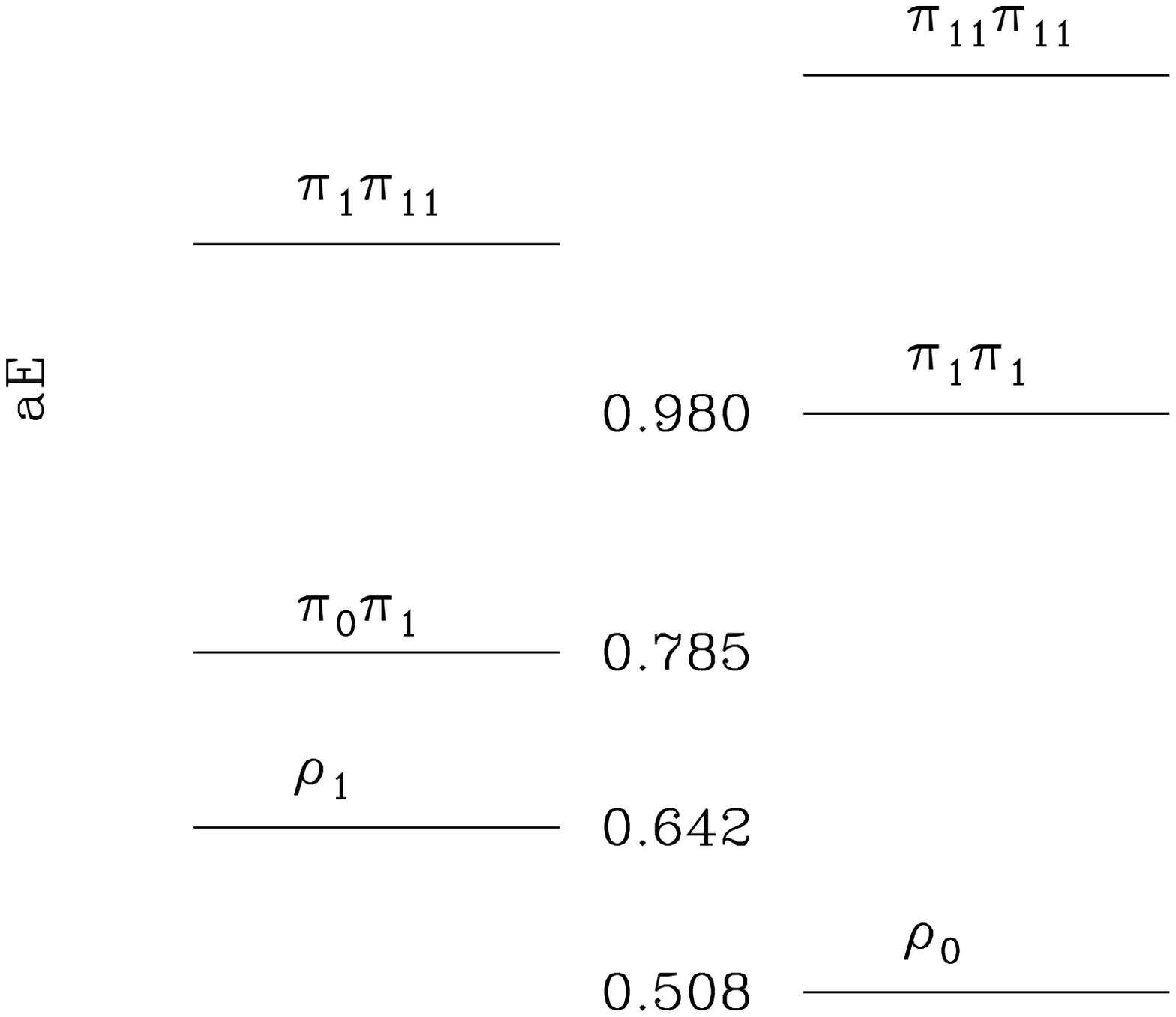}
 \vspace{-2.5cm}
  \caption{}
 \label{fig.rpp}
\end{figure}

\begin{figure}[th]

  \includegraphics[scale=0.4]{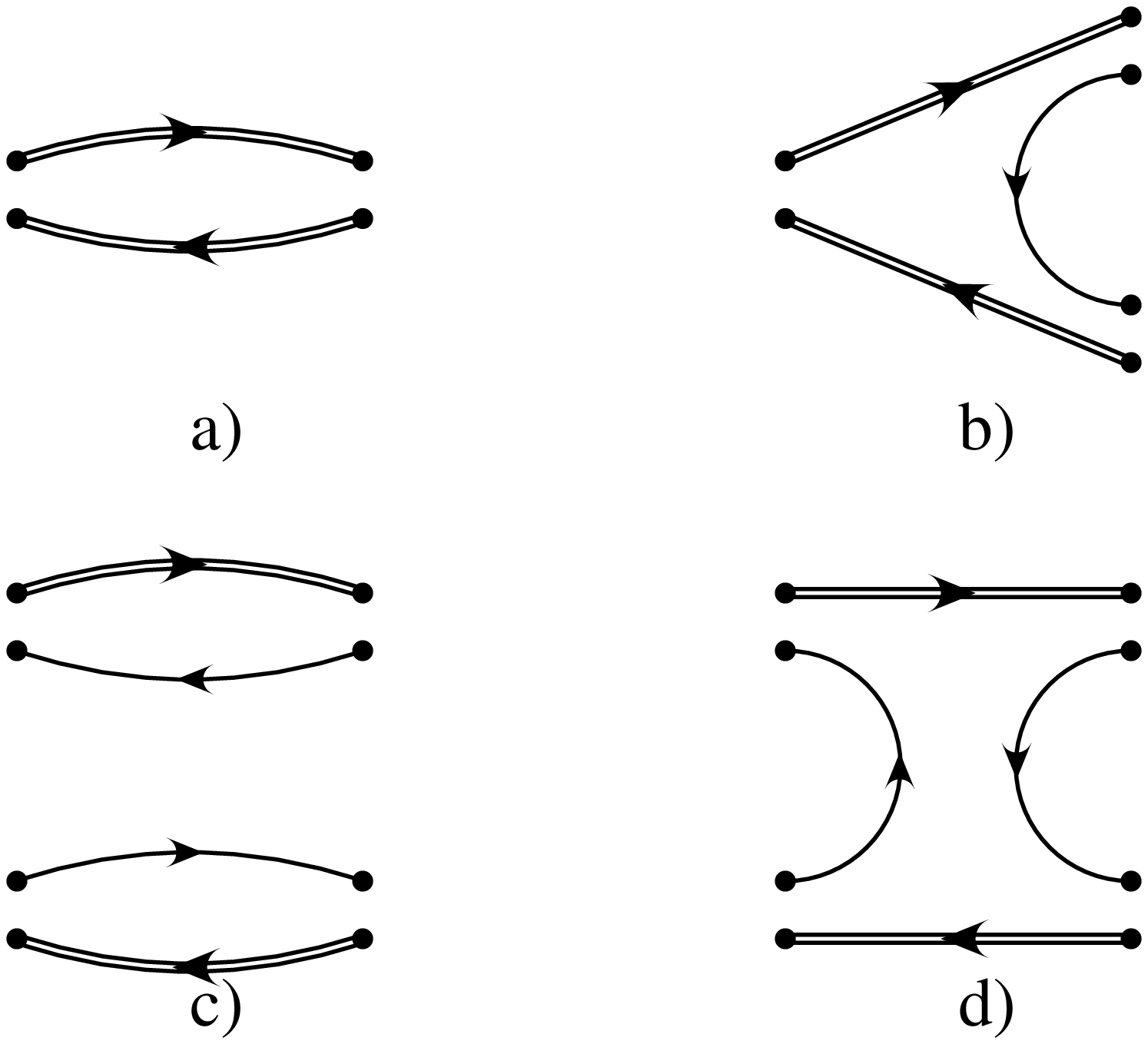}
\vspace{-1.0cm} 
 \caption{  }
 \label{fig.quark}
\end{figure}

\begin{figure}[th]
 \vspace{-2.5cm}
  \includegraphics[scale=0.3]{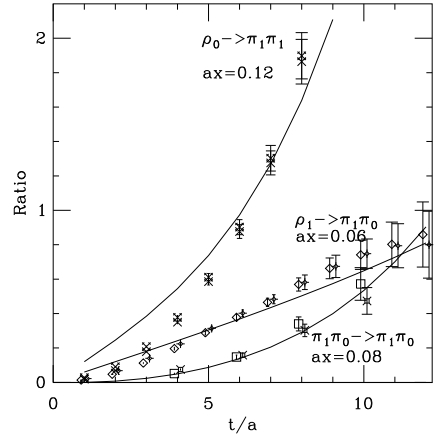}
 \vspace{-1.5cm} 
 \caption{Ratios of 3 and 4-point correlators.}
 \label{fig.xtall} 
 \end{figure}

 The situation we shall analyse is represented by the energy spectrum 
shown in fig.~\ref{fig.rpp}, here neglecting any interactions among the
states. We evaluate correlations between lattice operators creating both
a $\rho$ meson and a $\pi \pi$ state, using a stochastic  method to
evaluate the quark diagrams shown in fig.~\ref{fig.quark} from 20 gauge
configurations~\cite{ukqcd} (double the number  used in our initial
calculation) with $N_f=2$ sea quarks  of mass corresponding to about 2/3
of the strange-quark mass. 
  These correlations, normalised by the two point functions as
appropriate,  are illustrated in fig.~\ref{fig.xtall}.  Here the
off-diagonal case  (labelled $\rho_1 \to \pi_1 \pi_0$) shows the
important feature that it grows  approximately linearly with increasing
$t$. This  linear growth will only occur for on-shell
transitions~\cite{hdecay} and this is  essentially the case here.

 To extract an estimate of the transition amplitude, we introduce  a
parameter  $x=\langle \rho | \pi \pi \rangle$ where these states are 
normalised on the lattice (to unity).  If higher excited states are
neglected,  one can make a two state model (with basis states $\rho$ and
the lightest $\pi \pi$ state)  with this transition amplitude and
evaluate the contributions shown in fig.~\ref{fig.xtall}. Unlike weak
decays, there is no specific operator that causes the transition in
strong decays, hence the specific time (between 0 and $t$) of any
transition, such as $\rho \rightarrow \pi \pi$, is not known. In general
this makes it hard to control the contributions from excited states. As 
emphasized previously~\cite{hdecay}, these excited state contributions
can  be  avoided if the transition is approximately on-shell, when the 
linear dependence on $t$ of the off-diagonal transition amplitude (with
the slope $x$) is a unique  signature of this on-shell transition.

 Note that for the case  of $\rho_0 \to \pi_1 \pi_1$, the on-shell
condition is much less well satisfied but the relative momentum of the
pions in the centre of mass is twice as large as for $\rho_1 \to \pi_0
\pi_1$ and hence  $x$ should be approximately twice as large,
since for a P-wave decay there will be a momentum factor  in the
transition amplitude.

 A further check of the extraction of $x$ comes from the  box diagram, 
fig.~\ref{fig.quark}d,  which will have a contribution  behaving as $x^2
t^2/2$ arising from a $\rho$ intermediate state - see
fig.~\ref{fig.xtall}.

 For a  more quantitative estimate of $x$, one can suppress excited 
state contributions in analysing the data of fig.~\ref{fig.xtall}, as
discussed in ref.~\cite{rhodecay}.

 Note that the above methods could be used, in principle, even in the
quenched approximation.  They depend on assuming that $xa$ is fairly
weak, as we indeed find.

\section{ENERGY SHIFTS}

 A more rigorous approach is to focus on energy values. When two levels
are close (our on-shell condition), then they will mix and the resultant
energy  shifts give relevant information~\cite{luscher}.  Moreover we
can estimate these  energy shifts from our $x$-value which provides more
cross checks. These shifts can only be studied using dynamical fermions.

\begin{figure}[th]
 \vspace{-2.5cm}
  \includegraphics[scale=0.3]{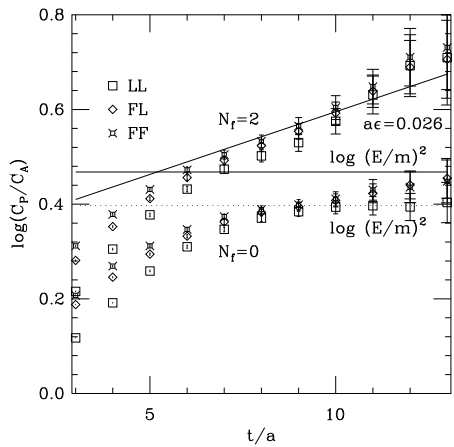}
 \vspace{-1.5cm}
 \caption{The ratio of Parallel to Antiparallel $\rho$ correlators.} 
 \label{fig.rhoratall}

\end{figure}

From a full variational analysis we obtain the energy shift of the
$\pi_1 \pi_0$ state (i.e. the un-binding energy), as needed in L\"uscher's
approach, as 0.02(2) upward which is consistent but not sufficiently
accurate to use. The energy shift of the $\rho_1$ state can however be
determined because of a lattice artifact. The $\rho$ with momentum 1 (in
lattice units of $2 \pi/L$) can have its spin aligned parallel to the
momentum axis (P) or perpendicular to it (A). Because the $\pi_0 \pi_1$
state has relative momentum along a lattice axis and the transition from
$\rho$ to $\pi \pi$ has orbital angular momentum L=1 (so a distribution
like $\cos \theta$), only the parallel state (P) can mix with this two
pion state. This mixing will not be present in the quenched
approximation, so this provides a direct opportunity to see the effect
of the two pion channel on the $\rho$ in unquenched studies. We do
indeed find such a mass splitting between the P and A orientations of 
the $\rho$  due to mixing for $N_f=2$ but not for $N_f=0$ as shown in
fig.~\ref{fig.rhoratall}. Moreover the magnitude of this energy shift
($0.026\pm 7$ in lattice units) is consistent with other determinations
of the transition strength.

\section{PHENOMENOLOGY}

 The basic assumption is that the transition from $\rho$ to $\pi \pi$ 
is given by an effective interaction with a finite spatial extent, this
is  usually summarised by an effective lagrangian where we normalise the
coupling  as $\bar{g}^2=\Gamma M E/k^3$ in terms of the decay width.
 Then, provided the lattice spatial size is big enough that the hadrons
are not distorted,  our lattice situation (where no decay occurs) can be
used  to determine $\bar{g}$ and this can then be used to predict decay
widths  when quark masses are varied, assuming that the coupling is
largely independent of the quark  masses.

 From our lattice studies, we deduce that $ax=0.06^{+2}_{-1}$ for 
$\rho_1 \to \pi_1 \pi_0$. Translating~\cite{rhodecay} this lattice
transition amplitude to the  continuum normalisation, gives
$\bar{g}=1.40^{+27}_{-23}$. Using the observed $\rho_1$ energy shift 
gives another estimate, namely $\bar{g}=1.56^{+21}_{-13}$.  Note that
our lattice values  would need to be extrapolated to light sea-quarks
and to the continuum limit  to allow all sources of systematic error to
be explored. Nevertheless, these  two values agree well with the
values extracted from decays of $\rho$, $K^*$ and $\phi$ mesons, namely
$\bar{g} \approx 1.5$.


Encouraged by the success of the calculation of the decay width of the
$\rho$ meson, we are starting to look at the decays of scalar mesons on
the lattice. This is important for studying the glue components of
singlet scalar mesons and to study $a_0(980)$ and $f_0(980)$ mesons
which are believed to have important 4 quark contributions.  There has
been one earlier calculation of the decay width of the $0^{++}$ glueball
from lattice QCD, but this has not been exploited by 
phenomenology~\cite{Close:2001zp}.

For  decays of singlet scalar mesons there are additional diagrams to
those in figure~\ref{fig.quark} from disconnected  graphs. The ``noise''
from the additional diagrams makes the  extraction of a signal hard. In
the real world the non-singlet $a_0(980)$ meson has the strong decays
$a_0 \rightarrow \pi \eta$ and  $a_0 \rightarrow K \overline{K}$. A
direct comparison with experiment will be non-trivial because the $\pi
\eta$ decay involves a noisy disconnected loop and  the $a_0 \rightarrow
K \overline{K}$ decay requires $N_f=2 +1$ sea quarks for a theoretically
clean calculation in this channel.

\end{document}